\documentclass[
 prd,
 amsmath,
 amssymb,
 reprint,
 twocolumn,
]{revtex4-1}

\usepackage{graphicx}
\usepackage{dcolumn}
\usepackage{bm}
\usepackage{braket}
\usepackage{color}
\usepackage{xcolor}
\usepackage{tabularx}
\usepackage{mathrsfs}
\usepackage{commath}
\usepackage{hyperref}
\usepackage{lineno}
\usepackage{float}
\usepackage[capitalise]{cleveref}

\hypersetup{
    colorlinks=true,
    linkcolor=blue,
    filecolor=magenta,      
    urlcolor=cyan,
    pdftitle={Overleaf Example},
    pdfpagemode=FullScreen,
    }

\begin{document}


\title{Guiding interferometer improvements with the frequency-dependent inspiral range} 



\author{Derek~Davis$^1$ and Elenna Capote$^{1,2}$}
\affiliation{$^1$LIGO, California Institute of Technology, Pasadena, CA 91125, USA}
\affiliation{$^2$LIGO Hanford Observatory, Richland, WA 99352, USA}
\date{\today} 

\begin{abstract}
The inspiral range is the most common metric for characterizing the performance of ground-based gravitational-wave interferometers. However, there is no clear formalism for working with frequency-dependent inspiral range quantities. We introduce a metric for the cumulative normalized range of a gravitational-wave interferometer, as well as methods to compare two separate noise curves. We show how this metric is a valuable tool for guiding the commissioning of these interferometers and provides increased clarity compared to other commonly used approaches. 
\end{abstract}

\maketitle 

\section{Introduction}

Ground-based gravitational-wave interferometers, such as the Advanced LIGO~\cite{aligo}, Advanced Virgo~\cite{avirgo}, and KAGRA~\cite{kagra} detectors now routinely detect signals from transient sources of gravitational waves. 
The past decade of observations has revolutionized our understanding of the universe, allowing us to probe the physics of matter at high densities~\cite{TheLIGOScientific:2017qsa}, discover new populations of stellar remnants~\cite{LIGOScientific:2020kqk}, and perform precision tests of general relativity~\cite{LIGOScientific:2020tif}.
Future improvements in this field will rely on continued improvements to current detectors~\cite{Aplus_design,PostO5,Flaminio:2020lqk} as well as the construction of next-generation facilities such as Cosmic Explorer~\cite{CE_horizon} and Einstein Telescope~\cite{ET_design}.

When evaluating the performance of these detectors for use in transient gravitational-wave astronomy, the primary figure of merit is the binary neutron star (BNS) inspiral range (which we will refer to simply as the \emph{inspiral range}), designed to roughly estimate the distance that a prototypical BNS signal could be detected~\cite{FinnChernoff:1993, LIGOScientific:2005gax, PSD:S5, Chen:2017wpg}. 
As a BNS inspiral sweeps through the majority of the most sensitive band of these detectors, this statistic condenses the broadband sensitivity of an interferometer into a single number. 
Almost every publication that discusses the sensitivity of current gravitational-wave detectors has used this metric as a primary figure-of-merit (e.g., Refs.~\cite{LIGOS2iul,aLIGO:2020wna,Capote:2024rmo}).

However, when investigating how to best improve a detector's sensitivity, it is often necessary to have frequency-dependent information in addition to this broadband figure of merit. 
For this reason, multiple approaches have been used to include aspects of the inspiral range in a frequency-dependent statistic, such as
the range integrand~\cite{sensemon_S3,Meadors:2014mon,Aplus_design} and
the cumulative inspiral range~\cite{Dwyer:2013dba}.
The quadrature nature of the inspiral range makes it challenging to compare the impact of different frequency bands to the overall sensitivity with these metrics. 
Furthermore, these metrics are not well suited to comparing the sensitivity of multiple detectors (or different time periods from the same detector), a common commissioning technique~\cite{alog:31442, alog:47209, alog:76768, vlog:62815}. 

To address these limitations, we introduce the
\emph{cumulative normalized range}, 
which is significantly easier to interpret compared to the previously listed techniques. 
The cumulative normalized range formalism also can naturally be extended to comparing the sensitivity of multiple detectors, resolving a common misconception in this calculation.
This cumulative normalized range formalism is more intuitive to interpret compared to previous methods and can be used in a wide variety of detector-focused investigations. 

This paper continues as follows. 
In \cref{sec:inspiral}, we outline the inspiral range as it is currently used in ground-based gravitational-wave astronomy. 
\cref{sec:cum_range} introduces the cumulative normalized range as the preferred quantity to evaluate the frequency-dependent sensitivity of gravitational-wave detectors, while \cref{sec:range_compare} shows how to apply this same formalism to compare the sensitivity of two interferometer sensitivity curves. 
We then show in \cref{sec:commis} how these techniques can be applied in practice by looking at a number of common examples in gravitational-wave detector investigations. 
We conclude in \cref{sec:conclu} discussing this formalism's limitations and provide several suggestions for other analysts that use this approach. 

\section{Derivation of the inspiral range} \label{sec:inspiral}

The signal-to-noise (SNR) of a gravitational-wave signal, $\rho$, is given by~\cite{LIGOScientific:2019hgc}
\begin{equation}
    \langle \rho \rangle= \sqrt{4 \int_{0}^{\infty} \frac{|\tilde{h}(f)|^2}{S_n(f)} df},
\end{equation}
where $S_n(f)$ represents the power spectral density of strain noise in a gravitational-wave interferometer. Inherent in this assumption is that the detector noise is stationary, and therefore the SNR can be totally determined by $S_n(f)$ and a waveform of the desired astrophysical system, $\tilde{h}(f)$.

We can approximate the waveform of a compact binary inspiral, $\tilde{h}(f)$, as~\cite{VIRGO:2012kov}
\begin{equation}
    \tilde{h}(f) =  \frac{1}{R}\left(\frac{5\pi }{24c^3}\right)^{1/2}(G\mathcal{M})^{5/6}(\pi f)^{-7/6} e^{i\Psi(f;M)},
\end{equation}
where $\mathcal{M}$ is the chirp mass~\cite{Cutler:1994ys} of the system and $\Psi(f;M)$ is some real function of frequency, parameterized by the total mass of the system, $M$. Solving for $R$ in terms of a SNR threshold, $\rho_0$, we find
\begin{equation}
    R = \frac{1}{\rho_0} \left( \frac{5\pi}{24c^3}\right)^{1/2} \frac{(G \mathcal{M})^{5/6} \pi^{-7/6}}{2.26} \sqrt{4 \int_{0}^{\infty} \frac{f^{-7/3}}{S_n(f)} df},
\label{eq:full_range}
\end{equation}
with an additional factor of $(2.26)^{-1}$ included to account for the average response of the detector across the entire sky~\cite{FinnChernoff:1993}. 
This quantity is generically referred to as the \emph{inspiral range} of the gravitational-wave detector~\cite{FinnChernoff:1993, LIGOScientific:2005gax, PSD:S5, Chen:2017wpg}.
For most range calculations, $\rho_0=8$ by convention. Commonly, the inspiral range is calculated with $\mathcal{M}\approx 1.218$\,M$_{\odot}$, corresponding to a system of two objects with masses 1.4\,M$_{\odot}$.
When this approximation for the compact binary inspiral waveform is used, \cref{eq:full_range} is also referred to as the ``sensemon range''~\cite{PSD:S5}.

For simplicity and generality in this paper, we can rewrite \cref{eq:full_range} as
\begin{equation}
\label{eq:range_simp}
    R = \sqrt{\int_{0}^{\infty} [r(f)]^2 df}.
\end{equation}
In practice, non-zero and finite integration bounds are used in \cref{eq:range_simp}. 
The frequency range where the detector data is properly calibrated guides the choice of frequency limits.   
The term $r(f)$ is often referred to as the \emph{range integrand}, with units of Mpc/$\sqrt{\mathrm{Hz}}$.

This simple form elucidates how the range is a parameter that should be added in quadrature in that it is the root of an integral (i.e., a sum) of squares. 
However, this functional form is also the source of numerous misunderstandings related to the range. 
One useful feature of the inspiral range is that it is proportional to the overall amplitude of noise in the interferometer; 
for example, if the amplitude of noise in an interferometer is halved at all frequencies ($\frac{1}{4}S_n(f)$ in power), the range will be doubled as 
\begin{equation}
    \sqrt{\int_{0}^{\infty} [2r(f)]^2 df} = 2R.
\end{equation} 
However, we can also see that the same linear relationship does not apply when adding two frequency bands of equal sensitivity. 
For example, if we fix
\begin{equation}
    \sqrt{\int_{0}^{f'} [r(f)]^2 df} = \sqrt{\int_{f'}^{\infty} [r(f)]^2 df} = \frac{1}{2}R_0, 
\end{equation} 
for a given $r(f)$, we find that 
\begin{align}
\begin{split}
    \sqrt{\int_{0}^{\infty} [r(f)]^2 df} &= \sqrt{\int_{0}^{f'} [r(f)]^2 df + \int_{f'}^{\infty} [r(f)]^2 df} \\
    &= \sqrt{\left(\frac{R_0}{2}\right)^2 + \left(\frac{R_0}{2}\right)^2} \\
    &= \frac{1}{\sqrt{2}}R_0 \neq R_0.
\end{split}
\end{align} 
Therefore, care must be taken when using \cref{eq:range_simp} with restricted frequency bounds to compare the contribution to the overall inspiral range from different frequency bands. 
The remainder of this work focuses on how to account for these properties of the inspiral range in a self-consistent manner when comparing the relative sensitivity between two frequency bands or two different spectra.

\section{Cumulative normalized range} \label{sec:cum_range}

It is convenient to focus on the range as a function of frequency, especially considering that upgrades in real detectors generally improve sensitivity in only parts of the detection band. 
Using the simplified \cref{eq:range_simp}, we can discuss an additional metric to allow the frequency-dependent investigation of the sensitivity of a gravitational-wave interferometer.

The cumulative range, roughly defined as the range of the detector if only frequencies below $f$ are considered, is commonly used to evaluate the frequency-dependent sensitivity of gravitational-wave detectors~\cite{alog:31442, alog:47209, alog:76768, vlog:62815}.
This cumulative range, $\hat{r}(f)$, is generally defined as~\cite{gwsumm:2.2.7}
\begin{equation}
    \hat{r}(f) = \sqrt{ \int_{0}^{f} [r(f')]^2 df' }.
    \label{eq:cum_range}
\end{equation}

However, as the range is calculated via quadrature sum, representing an incomplete sum is difficult to do in a self-consistent manner. 
To address this, we introduce the \emph{cumulative normalized range}, $\overline{r}(f)$, that resolves many of the limitations of classical cumulative range calculations.

\subsection{Derivation of the cumulative normalized range}

To be maximally useful as a commissioning tool, $\overline{r}(f)$ must obey the following principles:
\begin{itemize}
    \item \textbf{Linear function of sensitivity}. Specifically, we require that if $r(f_n) = r(f_m)$, then $\od{}{f}\overline{r}(f_n) = \od{}{f}\overline{r}(f_m)$. 
    \item \textbf{Follow expected limits}. The lower and upper limits of $\overline{r}(f)$ are well defined: $\overline{r}(0) = 0$ and $\overline{r}(\infty) = R$, where $R$ is in units of Mpc.
\end{itemize}

With these requirements in mind, $\overline{r}(f)$ must be of the form 
\begin{equation} 
    \overline{r}(f) = \int_{0}^{f} C[r(f')]^2 df',
\end{equation}
for some constant $C$ that has units of $\mathrm{Mpc}^{-1}$.
From the limiting behavior of this expression, we can also find that $C=1/R$, as $\int_{0}^{\infty} C[r(f')]^2 df' = R^2$. 
Therefore, $\overline{r}(f)$ can be written as
\begin{equation} \label{eq:r_sum}
    \overline{r}(f) = \int_{0}^{f} \frac{[r(f')]^2}{R} df',
\end{equation}
where $R$ is defined in \cref{eq:range_simp}.
As the measurability of many gravitational-wave parameters is proportional to $\rho^2$~\cite{LIGOScientific:2019hgc}, one can consider $\overline{r}(f)$ also to be proportional to the cumulative information gained about gravitational-wave event parameters with respect to frequency. 

Note that the normalized cumulative range is not equal to the cumulative range, i.e.,
\begin{equation}
    \overline{r}(f) \neq \hat{r}(f), 
\end{equation}
for $f\neq 0, \infty$. Therefore, $\overline{r}(f)$ differs from the range calculated with confined frequency bounds.

\subsection{Cumulative normalized range examples}

\begin{figure}[t]
    \centering
    \includegraphics[width=\columnwidth]{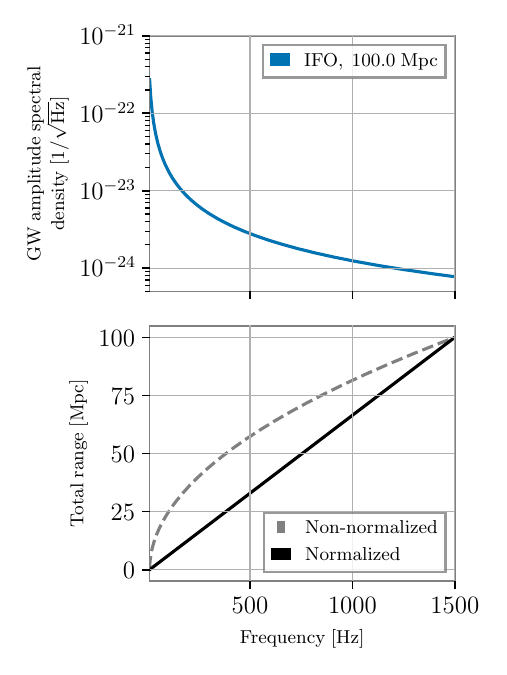}
    \caption{The top panel demonstrates a detector strain amplitude spectral density for the case in which the range integrand $r(f)$ is constant with frequency, $S_n(f) \propto f^{-7/3}$. The lower plot compares two methods of calculating the cumulative range of this curve for frequencies below $f$. The cumulative range method (gray dotted curve) misleadingly indicates that more range is accumulated at lower frequencies, while the cumulative normalized range method (black curve) demonstrates a linear accumulation of range.}
    \label{fig:constant_range}
\end{figure}

We can first show the utility of this expression by plotting both the cumulative range (\cref{eq:cum_range}) and cumulative normalized range (\cref{eq:r_sum}) for the contrived case where $S_n(f) \propto f^{-7/3}$, such that $r(f)$ is constant with frequency, as shown in the upper plot of \cref{fig:constant_range}. 
In this scenario, we expect the cumulative normalized range to be a straight line. 
This is indeed the case, as is shown by the black curve in the lower plot of \cref{fig:constant_range}. However, the cumulative (non-normalized) range shown in the gray dotted curve indicates that more range is accumulated at lower frequencies than at higher frequencies. The opposite would be demonstrated if the integration was performed in reverse, i.e., from $\infty$ to $f$. The overall effect is that the cumulative range is misleading about what fraction of the total range is contributed by each frequency within the detection band.

\begin{figure}[t]
    \centering
    \includegraphics[width=\columnwidth]{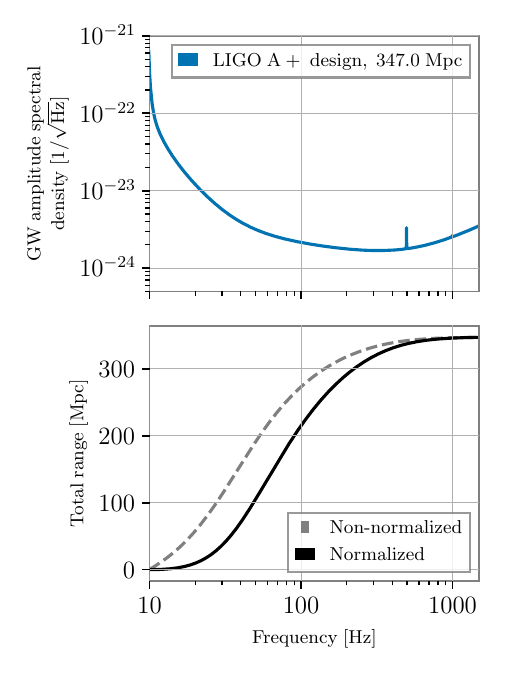}
    \caption{Example of the cumulative range functions for the upgrade to Advanced LIGO (A+) design curve (top). The non-normalized cumulative range attributes significantly more range gain from the sensitivity at low frequency, nearly 30\% below 30\,Hz. However, the cumulative normalized range calculation demonstrates that actually 30\% of the range will accumulate below 50\,Hz. Note the logarithmic x-axis.
    }
    \label{fig:aligo_example}
\end{figure}

We also apply this method to the upgrade of Advanced LIGO (known as A+) design sensitivity curve~\cite{Aplus_design} depicted in the upper plot of \cref{fig:aligo_example} with a total range of 347\,Mpc. The lower plot also compares the cumulative and cumulative normalized range methods. The cumulative range calculation indicates that more sensitivity will be gained at low frequency with nearly 30\% of the total A+ sensitivity between 10--30\,Hz. The cumulative normalized range calculation shows that this is not actually the case. Conversely, the cumulative range understates the importance of higher frequency contributions to the range.

\section{Cumulative range comparisons} \label{sec:range_compare}

When comparing the cumulative range between two different spectra, additional care must be taken when renormalizing this quantity. 
As the cumulative range is a quadrature sum, the difference in cumulative sensitivity can be characterized as $R_i \overline{r}_i(f) - R_j \overline{r}_j(f)$.
However, this term is difficult to work with as it is in units of Mpc$^2$.
We, therefore, must also change how we calculate the cumulative range difference.

\subsection{Derivation of the normalized range difference}

To derive the \emph{cumulative normalized range difference}, $\overline{r}_{ij}(f)$ we follow similar constraints as the cumulative normalized range: 
\begin{itemize}
    \item The normalized function should obey the requirement that if $[r_i(f_n)]^2 - [r_j(f_n)]^2 = [r_i(f_m)]^2 - [r_j(f_m)]^2$, then $\od{}{f}\overline{r}_{ij}(f_n) = \od{}{f}\overline{r}_{ij}(f_m)$. This points to a normalization of the form $C (R_i \overline{r}_i(f) - R_j \overline{r}_j(f))$ for some constant $C$.
    \item The normalized function should have the expected behavior in the low and high-frequency limits --- namely  0 at the low-frequency limit, and then $R_i - R_j$ (the difference in the ranges) at the high-frequency limit.
\end{itemize}

We then solve for $C$ following the second constraint:
\begin{equation}
\begin{split}
R_i - R_j &= \lim_{f\to\infty} C ( R_i \overline{r}_i(f) - R_j \overline{r}_j(f)) \\
R_i - R_j &= C ( R_i^2 - R_j^2) \\
C &= 1/ ( R_i + R_j) \\
\end{split}
\end{equation}

Therefore, the proper normalization of the difference in cumulative ranges is 
\begin{equation} \label{eq:r_diff}
\overline{r}_{ij}(f) = \frac{R_i \overline{r}_i(f) - R_j \overline{r}_j(f)}{R_i + R_j},
\end{equation}

We can view this expression in contrast with 
\begin{equation}
    \Delta R_{ij}(f) = \sqrt{ \int_{0}^{f} [r_i(f')]^2 df' } - \sqrt{ \int_{0}^{f} [r_j(f')]^2 df' }.
\label{eq:delta_R}
\end{equation}
This term, $\Delta R_{ij}(f)$, has been referred to as the \emph{cumulative range difference}. It is perhaps the most straightforward approach to comparing the range from two different noise curves, but it provides highly misleading information, as we will see.  

\subsection{Normalized range difference example}

\begin{figure}[t]
    \centering
    \includegraphics[width=\columnwidth]{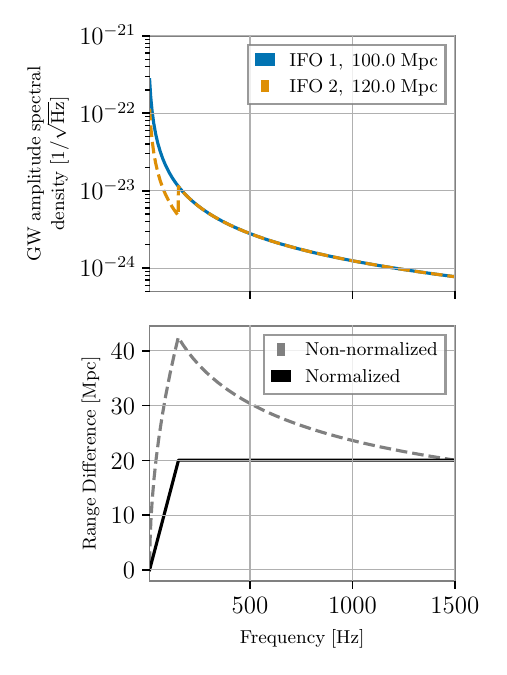}
    \caption{The top plot compares two strain amplitude spectral densities, that of IFO\,1 with 100\,Mpc of range (blue solid curve) and IFO\,2 with 120\,Mpc of range (orange dotted curve). The increase in range of IFO\,2 has occurred within the 10--150\,Hz band, and the two interferometers have the same sensitivity above 150\,Hz. The lower plot compares the cumulative range difference with the non-normalized (grey dotted curve) and normalized methods (black solid curve). By simply subtracting the two cumulative ranges, it appears that IFO\,2 gains much more sensitivity below 150\,Hz relative to IFO\,1, and then loses much of that sensitivity above 150\,Hz to maintain the correct 20\,Mpc range difference at the upper limit. By applying \cref{eq:r_diff}, the sensitivity increase in IFO\,2 from 10--150\,Hz is evident, as well as the equivalency in the sensitivity between the two interferometers above 150\,Hz.}
    \label{fig:range_diff}
\end{figure}

\begin{figure*}[tb] 
    \centering
    \includegraphics[width=\textwidth]{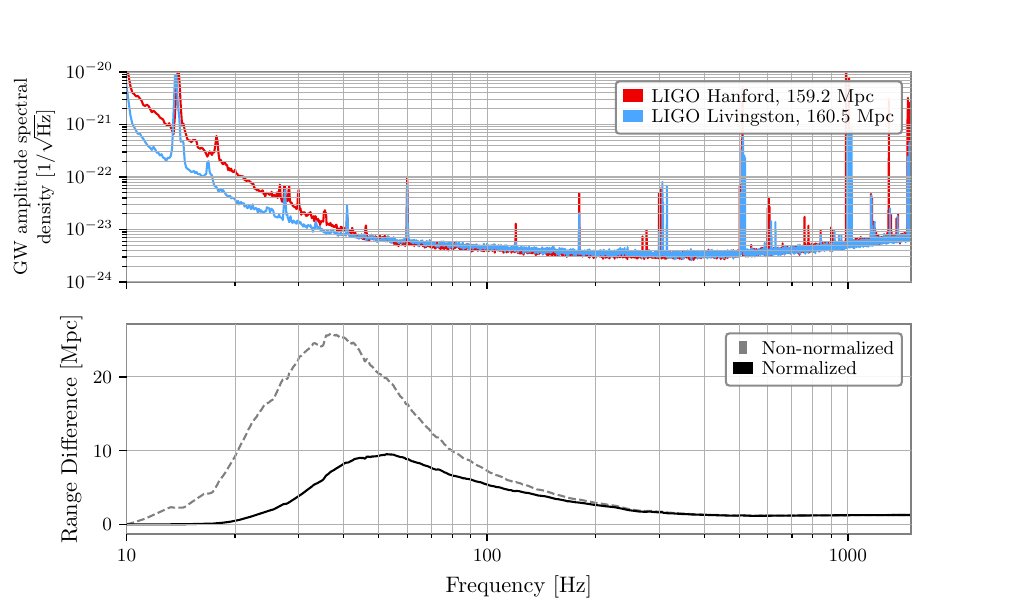}
    \caption{A comparison of the LIGO Livingston and LIGO Hanford detector sensitivities (upper) and their corresponding range difference (lower), calculated using the non-normalized and normalized methods. The strain sensitivities here correspond to the strain sensitivities of each detector during the first part of the fourth observing run, as in~\cite{Capote:2024rmo}, chosen such that the total range is similar. However, there is a stark difference in the strain curves between 10--40\,Hz. The non-normalized range difference indicates the loss in range at low frequency is large, implying that 10\,Mpc of sensitivity is lost in the Hanford sensitivity from 10--20\,Hz alone. However, the normalized range quantity demonstrates that the range difference in this band is negligible.}
    \label{fig:O4a_ligo}
\end{figure*}

You can see application of Equations~\ref{eq:r_sum} and~\ref{eq:r_diff} in Figure~\ref{fig:range_diff}. Figure~\ref{fig:range_diff} represents a contrived situation in which two interferometer sensitivity curves (``IFO\,1'' and ```IFO\,2'') have the same sensitivity above 150\,Hz, but IFO\,2 is more sensitive in the band of 10--150\,Hz, such that the total range of IFO\,2 is 20\,Mpc greater than IFO\,1. A straightforward subtraction of the two cumulative ranges following \cref{eq:delta_R} indicates that the sensitivity gained in the second IFO from 10--150\,Hz is much higher than 20\,Mpc. As such, the difference between the two cumulative ranges must then decrease over the rest of the band so that the true difference in the range is achieved at the upper limit. This makes it appear that the second interferometer has gained significantly more sensitivity from 10--150\,Hz, and then lost most of that sensitivity from 150--1500\,Hz. By applying the more appropriate normalized cumulative range, \cref{eq:r_diff}, the increase from 10--150\,Hz and the equivalency in the two sensitivities above 150\,Hz are both evident.

\section{Use as a commissioning tool} \label{sec:commis}

It is a common practice in detector commissioning to compare two ranges, such as those between two detectors, or before and after detector upgrades. This is most beneficially done in a frequency-dependent way; most detector upgrades only affect certain frequency bands within the full sensitivity of the interferometer. It is also possible for a detector upgrade to increase the sensitivity in some bands and worsen it in others. Careful consideration of the full, frequency-dependent effect of detector changes must be made when evaluating the benefit of commissioning upgrades on the inspiral range.

\cref{fig:O4a_ligo} provides an example detector sensitivity comparison between the strain sensitivities of the LIGO Hanford (red) and LIGO Livingston detectors (blue) in the fourth observing run (O4). These strain sensitivities were chosen to demonstrate a time when the two detectors had a similar inspiral range. A significant difference in the two strain curves is evident from 10--40\,Hz. However, the overall range of these two curves is nearly the same, as indicated in the legend. The lower plot of \cref{fig:O4a_ligo} better captures where the LIGO Livingston detector achieves more or less range relative to LIGO Hanford. Without normalization, it appears that LIGO Hanford has 10\,Mpc less range from 10--20\,Hz alone, but the normalized method indicates that the contribution to the overall difference in range in this band is almost negligible.

\begin{figure*}[tb]
    \centering
    \includegraphics[width=\textwidth]{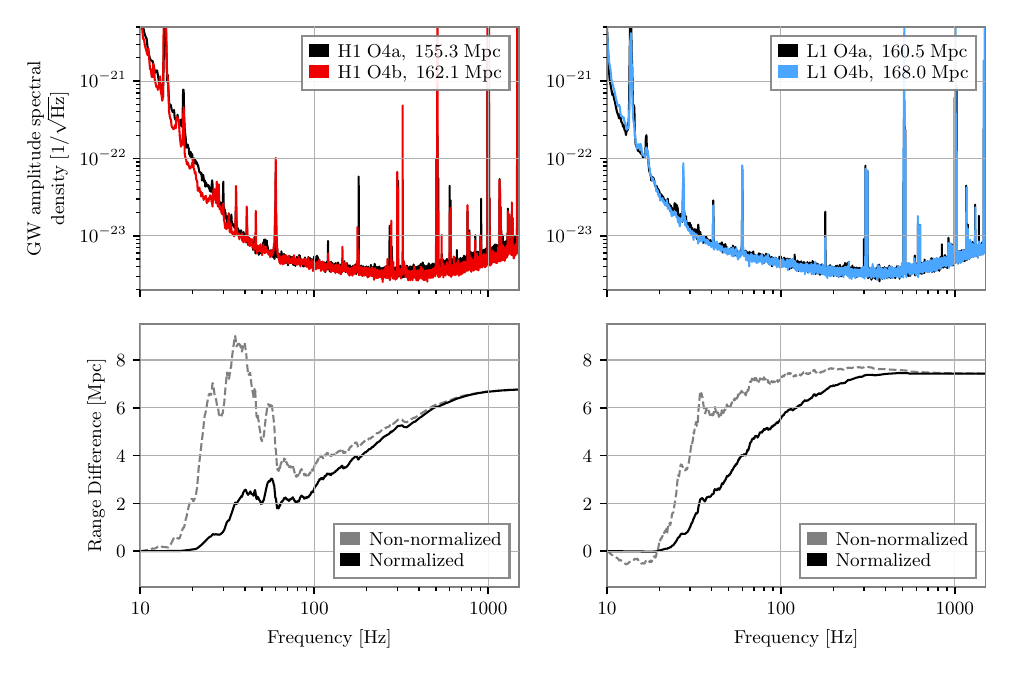}
    \caption{Comparisons of the sensitivity and range change at the LIGO Hanford (H1, left) and LIGO Livingston (L1, right) detectors between the first and second part of the fourth observing run (O4a and O4b). The top plots compare typical O4a and O4b sensitivities, with the total range indicated in the legend. 
    The bottom plots compare the normalized and non-normalized range differences for the two curves in the top plots. The normalized range difference demonstrates that the sensitivity improvements led to a gain in a range nearly everywhere, while the non-normalized range difference misleadingly indicates some range was lost during the upgrades.
    }
    \label{fig:o4}
\end{figure*}

Furthermore, the normalized range difference demonstrates that the seemingly minor improvement in the LIGO Hanford sensitivity from 50--200\,Hz relative to LIGO Livingston is nearly equivalent to the range lost at lower frequencies where the difference in sensitivity is more pronounced. This qualitative behavior is also evident in the non-normalized range difference. However, the inflection point of the range difference (which should occur when the LIGO Hanford sensitivity crosses the LIGO Livingston sensitivity) is only at the correct frequency in the normalized range difference. In fact, the extreme overshoot in range gain in the non-normalized difference artificially shifts the inflection point lower.

These characteristics of the two detector sensitivities can be understood by the different challenges faced by each detector during O4. 
The LIGO Hanford Observatory faced additional unknown technical noise from 10--40\,Hz, while the LIGO Livingston Observatory was challenged by a broadband unknown noise around 100\,Hz that limited the benefits of higher operating power and increased quantum squeezing~\cite{Capote:2024rmo}. 
Efforts during a commissioning break between the two parts of the fourth observing run (O4a and O4b) targeted some of these noise challenges. \cref{fig:o4} compares the sensitivity changes at each detector from O4a to O4b, using more representative sensitivity curves from each run part. 
Both detectors benefited from improvements in the injected quantum squeezing as well as reductions in technical noise that limited the low-frequency performance~\cite{Capote:2024rmo}.
The normalized range difference demonstrates that the sensitivity of both detectors increased across most of the band from O4a to O4b, with a few small features in the LIGO Hanford sensitivity that caused minor range reductions. 
Two drawbacks of the non-normalized difference method are shown: LIGO Hanford appears to lose range from 40--70\,Hz, when in fact the sensitivity did not change appreciably in that band, and LIGO Livingston appears to lose range from 10--20\,Hz, even though the change in sensitivity in that band has no effect on the range.

\begin{figure}[tb]
    \centering
    \includegraphics[width=\columnwidth]{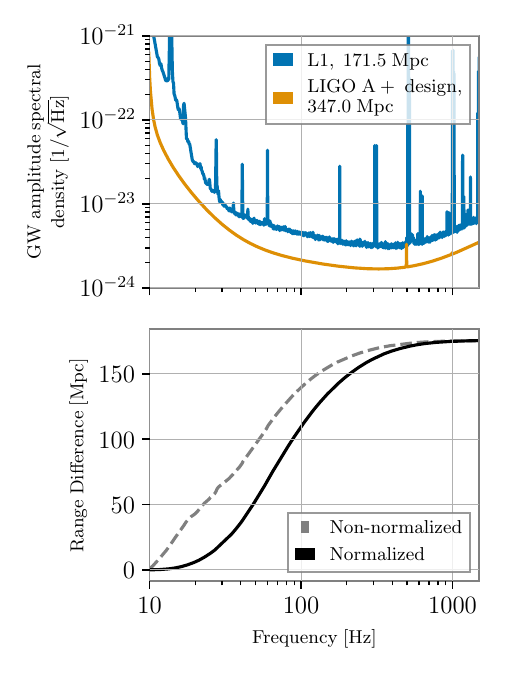}
    \caption{The top plot shows the best-achieved LIGO sensitivity to date, as measured at the LIGO Livingston Observatory (L1) during the second part of the fourth observing run. This sensitivity is compared with the design sensitivity for the upgrade to Advanced LIGO (A+). The bottom plot applies the range difference calculations using the non-normalized and normalized methods. The normalized method indicates that the main contribution to the range increase will be due to improvements in the 20--500\,Hz band. However, improvements to the sensitivity outside of this frequency band are still beneficial for gravitational-wave astrophysics, indicating the limitations of the inspiral range as a detector sensitivity metric.}
    \label{fig:design}
\end{figure}

Finally, an exemplar LIGO Livingston sensitivity curve is compared with the LIGO A+ design sensitivity curve in \cref{fig:design}. 
The first part of the A+ upgrade program was implemented during the preparation for O4~\cite{Capote:2024rmo}, and the remainder will be completed by the end of the fifth observing run (O5)~\cite{Capote:2024rmo,Aasi:2013wya}. 
Reaching the design sensitivity would double the best-achieved LIGO detector range and require improvements in the sensitivity everywhere in the band. 
The normalized range difference demonstrates that the main contribution to the range increase will occur from 20--500\,Hz. This fact highlights one limitation of the inspiral range as a detector sensitivity metric. 
The gain in sensitivity upon achieving design performance from 10--20\,Hz would greatly improve the scope of astrophysical observations that can be done with the LIGO detectors. 
For example, the improved sensitivity from 10--20\,Hz would probe higher--mass compact binary systems, particularly those about 100\,M$_{\odot}$~\cite{LIGOScientific:2021tfm}. 
Similarly, the improvements above 500\,Hz would benefit the study of high-frequency gravitational-wave astrophysics, such as the detection of a binary neutron star post-merger signal~\cite{Clark:2015zxa}. 
Therefore, \cref{fig:design} is a reminder of the limitations of the inspiral range as a metric to evaluate the performance of gravitational-wave detectors for all science cases.

\section{Conclusions} \label{sec:conclu}

By resolving many of the misconceptions regarding the inspiral range, the cumulative normalized range formalism introduced in this work provides a new way to understand the frequency bands that limit the performance of current and future gravitational-wave detectors. 
As current facilities approach their design sensitivities, the additional gains that can be realized will become much more incremental and targeted. 
The cumulative normalized range will be one of the important tools available to guide these detector improvements and ensure that gravitational-wave detectors can reach their full potential. 

The challenges discussed in this work become more pronounced as additional astrophysical features are added to the inspiral range calculation~\cite{Chen:2017wpg}. 
Such extensions to the inspiral range can improve the usefulness of this quantity for astrophysical projections, but make it more difficult to use as a tool for comparing the performance of gravitational-wave detectors in a self-consistent manner. 
For example, the inspiral range has historically been reported as the luminosity distance, as this ensures that the amplitude spectral density of the detector is linear with respect to the reported range. 
If, instead, the comoving distance is reported, this linear relationship is lost, introducing another non-commutative feature in the inspiral range calculation beyond the quadrature sum.
Including even more cosmological effects, such as the redshifting of the observed gravitational-wave signal, adds even more complications. 
The formalism in this paper does not address the additional challenges that the use of these cosmological effects would induce but could still be applied in these cases as long as the appropriate form of $r(f)$ is chosen. 

It should also be noted that there are other sensitivity metrics~\cite{Chen:2017wpg, Sutton:2013ooa, Gupta:2023lga} for gravitational-wave detectors that are focused on a variety of scientific observations.
For example, the measurability of many event parameters, such as the spin parameters of a detected signal, is not proportional to $f^{-7/3}$~\cite{Damour:2012yf}; these measurements are more strongly dependent on the low-frequency sensitivity than the inspiral range. 
There are also many expected sources of gravitational waves, such as the BNS post-merger~\cite{Clark:2015zxa} or magnetar bursts~\cite{LIGOScientific:2022sts}, at frequencies higher than included in the inspiral range. 
Many signals are not broadband, such as continuous wave sources~\cite{Riles:2022wwz}; hence, the sensitivity to these sources is already represented by the detector's sensitivity at the frequency of interest. 
Finally, the data quality can strongly impact the detectability and measurability of gravitational-wave signals and their properties~\cite{Davis:2022dnd,Capote:2024mqe}.
As no single metric can encompass all science cases, we strongly support using a broad array of different sensitivity metrics when evaluating the performance of current and future detectors~.

\section*{Acknowledgements} \label{sec:acknow}
We thank Anamaria Effler, Sheila Dwyer, Oli Patane, Gabriele Vajente, Ian MacMillan, and Edgard Bonilla for helpful discussions regarding the use of inspiral range and Jennie Wright for their comments during the internal review of this paper.
DD and EC are supported by the NSF through the LIGO Laboratory.
This material is based upon work supported by NSF’s LIGO Laboratory 
which is a major facility fully funded by the 
National Science Foundation.
LIGO was constructed by the California Institute of Technology 
and Massachusetts Institute of Technology with funding from 
the National Science Foundation, 
and operates under cooperative agreement PHY-2309200.
The authors are grateful
for computational resources provided by the LIGO Laboratory and supported by National Science Foundation
Grants PHY-0757058 and PHY-0823459.
This work carries LIGO document number P2500021.

\bibliography{main.bbl}

\end{document}